\documentclass[aps,floats,twocolumn,prl,nofootinbib]{revtex4-2}
\usepackage{amsfonts,amsmath,amssymb,ascmac,bm,tensor}
\usepackage{fnpct} 
\usepackage{comment}
\usepackage{ifpdf}
\usepackage{slashed}
\usepackage{color}
\usepackage[mathscr]{eucal}
\usepackage[utf8]{inputenc}
\usepackage{physics}
\usepackage{cancel}
\usepackage{subcaption}
\usepackage{soul}
\usepackage{simpler-wick}
\usepackage[breaklinks=true]{hyperref}
\def\ddd{\mathrm{d}}

\def\r{\right}

\begin{document}
\title{Fixing the Renormalization of Inflationary Loops via Ward Identities}
\author{Cheng-Jun Fang$^{1,2}$}
\email{fangchengjun@itp.ac.cn}
\author{Zong-Kuan Guo$^{1,2,3}$}
\email{guozk@itp.ac.cn}

    \affiliation{$^1$Institute of Theoretical Physics, Chinese Academy of Sciences, P.O. Box 2735, Beijing 100190, China}
    \affiliation{$^2$School of Physical Sciences, University of Chinese Academy of Sciences, No.19A Yuquan Road, Beijing 100049, China}
    \affiliation{$^3$School of Fundamental Physics and Mathematical Sciences, Hangzhou Institute for Advanced Study, University of Chinese Academy of Sciences, Hangzhou 310024, China}

\begin{abstract}
Evaluating quantum loop corrections to curvature perturbations in non-attractor inflation presents theoretical ambiguities. A crucial aspect of this challenge lies in the unconstrained finite contributions in renormalization counterterms and regularization scheme dependence. In this work, we derive exact Ward identities via the path integral formalism based on the large gauge symmetry of the background-perturbation split. These identities are shown to impose strict, model-independent constraints on the renormalization procedure. Provided the ultraviolet completion respects this symmetry, the Ward identities non-perturbatively govern the infrared evolution of the power spectrum. This symmetry-based framework offers a systematic resolution to recent theoretical discrepancies concerning one-loop corrections in ultra-slow-roll inflation.
\end{abstract}

\maketitle

\section{Introduction}
Inflation at the very beginning of the Universe \cite{Guth:1980zm,Starobinsky:1980te} is the leading paradigm to explain primordial perturbations. While large-scale observations \cite{Planck:2018jri,Planck:2019kim,ACT:2025fju,BOSS:2016wmc,DESI:2024mwx,DESI:2024hhd} perfectly match slow-roll (SR) predictions \cite{Starobinsky:1982ee,Mukhanov:1990me,MukhanovFeldmanBrandenberger1992}, indications from primordial black holes \cite{Green:2020jor,Carr:2020xqk,KAGRA:2021duu,DeLuca:2020agl} and the stochastic gravitational wave background \cite{NANOGrav:2023hde,NANOGrav:2023gor,Zic:2023gta,Reardon:2023gzh,EPTA:2023fyk,EPTA:2023sfo,Xu:2023wog} suggest significant small-scale amplification of the primordial power spectrum. Consequently, non-slow-roll models, such as ultra-slow-roll (USR) \cite{Namjoo:2012aa,Garcia-Bellido:2017mdw,Germani:2017bcs,Motohashi:2017kbs,Xu:2019bdp,Fu:2019vqc} or parametric resonance \cite{Cai:2018tuh,Cai:2019jah,Cai:2019bmk}, have been widely proposed.

The inherent amplification feature of these models, however, implies that they are likely to exhibit strong nonlinear effects, which have been shown in both theoretical and numerical analyses \cite{Inomata:2022yte,Caravano:2024tlp,Kristiano:2022maq,Kristiano:2023scm,Cheng:2021lif}. This requires us to treat loop-order corrections with extreme caution. Unfortunately, loop calculations in non-attractor inflation suffer from many fundamental difficulties caused by complex background dynamics, which have directly led to numerous controversies in recent years. Many works have introduced a variety of approaches \cite{Riotto:2023hoz, Choudhury:2023vuj, Choudhury:2023jlt, Choudhury:2023rks, Firouzjahi:2023aum, Motohashi:2023syh, Firouzjahi:2023ahg, Franciolini:2023agm, Cheng:2023ikq, Iacconi:2023ggt, Maity:2023qzw, Firouzjahi:2023bkt, Ballesteros:2024zdp, Kong:2024lac, Kristiano:2024ngc, kristianoComparingSharpSmooth2024, Fumagalli:2023zzl, Iacconi:2026uzo, Firouzjahi:2025ihn,Li:2026vrn, Cruces:2026qvl}, yet a full consensus on the outcome of loop corrections has still not been reached.

Recently, many studies have pointed out that regularization and renormalization are decisive steps in full loop diagram calculations \cite{Sheikhahmadi:2024peu, Inomata:2025bqw, Inomata:2025pqa, Braglia:2025cee, Braglia:2025qrb, Braglia:2026fle, Ema:2026dop, Ballesteros:2025nhz, Kristiano:2025ajj}, while some earlier works have postponed this part of the discussion. It has been argued in the literature that different regularization schemes can introduce different finite parts into the final loop results \cite{Kristiano:2025ajj}. Meanwhile, counterterms are inevitably required throughout the renormalization procedure, and the resulting finite physical predictions depend substantially on the finite contributions contained in these counterterms \cite{Inomata:2025pqa,Braglia:2025cee, Braglia:2025qrb, Braglia:2026fle, Ema:2026dop, Ballesteros:2025nhz}. This introduces inherent arbitrariness into the theoretical outcomes, thereby challenging the predictive power of the theory. 

This is quite a general problem in quantum field theory, and it has already been resolved in many branches of field theory. In gauge theory, if we adopt a symmetry-preserving regularization scheme \cite{tHooft:1972tcz}, such as dimensional regularization (dim-reg), or fine-tune the counterterm coefficients by virtue of Ward identities \cite{Peskin:1995ev, Kluberg-Stern:1974iel, Chen:2016nrs}, we can uniquely determine the renormalized loop results and ensure that they satisfy the symmetry requirements. Also, in de Sitter or SR attractor inflationary spacetimes, it has been pointed out in the literature that spacetime symmetries can forbid the late-time divergence of the renormalized power spectrum \cite{Senatore:2009cf,Assassi:2012et, Pimentel:2012tw, Gorbenko:2019rza, Farren:2026hao}. Notably, many recent works have revealed a profound connection between the evolution of the power spectrum and large gauge symmetries in non-attractor inflation. Several of these studies employ tree-level Maldacena consistency relations \cite{Maldacena:2002vr}, as well as symmetry constraints on four-point interactions and counterterms \cite{Inomata:2024lud,Inomata:2025bqw,Inomata:2025pqa,Tada:2023rgp,Fumagalli:2024jzz,Ema:2026dop,Kawaguchi:2024rsv,Fang:2025vhi}.  
Ref.\cite{Fang:2025kgf, Prokopec:2025jrd} further points out that Ward identities can impose non-perturbative constraints directly on the evolution of the power spectrum.

This strongly suggests that, within non-attractor inflation, one should likewise be able to develop an approach to eliminate the ambiguities in renormalization by means of symmetry. We find that, just as in gauge theories, Ward identities can also play a decisive role in this context. Even when facing issues such as an initial vacuum that breaks symmetries \cite{Hinterbichler:2013dpa, Assassi:2012zq, Kundu:2014gxa,Kundu:2015xta,Goldberger:2013rsa} and the violation of equal-time consistency relations in non-slow-roll scenarios \cite{Bravo:2017wyw,Cai:2018dkf}, the improved approach based on unequal-time Ward identities \cite{Hui:2018cag,Fang:2025kgf} can still fix the arbitrary terms arising in the renormalization procedure, yielding well-defined physical results. 

In this paper, we first review the loop diagram calculations for non-attractor inflation. We then extract the divergences of loop integrals and discuss the corresponding choice of counterterms. Finally, we introduce symmetries and Ward identities, and elaborate in detail how they fix the renormalized results.

\section{Revisiting One-Loop Diagrams}
To demonstrate the physical mechanism underlying loop corrections more clearly, we consider a relatively simple but highly representative scenario. We focus on a single-field inflation model within the spatially flat gauge. The evolution of the background stage is driven by a renormalized potential function $V_r(\phi)$, which is specifically constructed to generate an inflationary trajectory featuring an intermediate phase between two SR periods. 

We assume that the first SR parameter $\epsilon$ remains small throughout the entire inflationary process. This assumption is applicable to several well-motivated scenarios, including USR and parametric resonance models~\cite{Inomata:2024lud,Inomata:2022yte,Cai:2019bmk}. With this setup, the lapse and shift functions in the ADM formalism are heavily suppressed by $\epsilon$, thus fully justifying the decoupling limit. Under this setting, the complete Lagrangian of the model can be expressed as
\begin{equation}S = \int \mathrm{d}\eta \,\mathrm{d}^3 x \, a^4 \left[ -\frac{1}{2} \partial^\mu \phi \partial_\mu \phi - V_r(\phi) - \mathcal{L}_c \right].\end{equation}
The counterterms are introduced in a general form $\mathcal{L}_c$ to avoid any a priori assumptions about the ultraviolet (UV) structure of the theory. While tree-level predictions of single-field inflation models align perfectly with observational requirements, counterterms naturally arise at the loop order and reflect the properties of the 'bare' theory at extremely high cutoff scales. Therefore, we will progressively determine their explicit form in the subsequent calculations.

We can always separate a classical background $\bar{\phi}$ that satisfies the following equation of motion: $\bar{\phi}'' + 2\mathcal{H}\bar{\phi}' + a^2 V_r^{(1)}(\bar{\phi}) = 0$.
Defining the field perturbation as $\delta\phi \equiv \phi - \bar{\phi}$, we expand the Lagrangian with respect to these perturbations to extract the action for $\delta\phi$:
\begin{equation}
\begin{aligned}
S_{\delta\phi} =& \int \mathrm{d}\eta \,\mathrm{d}^3 x \, \frac{a^2}{2} \left[ (\delta \phi')^2 - (\partial_i \delta \phi)^2 \right] \\
-& a^4 \sum_{n=2}^{\infty} \frac{1}{n!} V_r^{(n)}(\bar{\phi}) \delta \phi^n - \mathcal{L}_c.
\end{aligned}
\end{equation}

To evaluate the one-loop order corrections to the power spectrum induced by the renormalized potential, we utilize the in-in formalism. For the issues investigated in this paper, the one-point and two-point correlation functions are of primary importance. 
The one-point correlation function $\langle\delta\phi\rangle$, often referred to as the backreaction term, can be represented by the diagram in Fig.~\ref{fig:fig1}(a). The loop corrections to the two-point correlation function are more complex, consisting of four distinct contributions. One of these, denoted as $P_{1\mathrm{vx}}$, involves a single vertex and corresponds to Fig.~\ref{fig:fig1}(b). Another single-vertex contribution appears with a tadpole topology, graphically represented in Fig.~\ref{fig:fig1}(c), which we call $P_{\mathrm{tad}}$. 
\begin{figure}[h]
  \centering
  \begin{minipage}[c]{0.25\linewidth}
    \centering
    \begin{tikzpicture}[line width=1pt,baseline=(v.base),scale=0.7]
      \coordinate (i) at (-1.2,0);
      \coordinate (v) at (0,0);
      \coordinate (o) at (1.2,0);
      \coordinate (b) at (0,-0.6);
      \def\r{0.4}
      \coordinate (c) at (0,-1);
      \draw (i) -- (o);
      \draw (c) circle (\r);
      \draw[thick] (b) -- (v) node[midway, sloped, allow upside down]{$>$};
    \end{tikzpicture}
    \caption*{(a) $\langle\delta\phi\rangle$}
    \label{fig:one-point}
  \end{minipage}%
  \hspace{0.04\linewidth}
  \begin{minipage}[c]{0.25\linewidth}
    \centering
     \begin{tikzpicture}[line width=1pt,baseline=(v.base),scale=0.7]
      \coordinate (i) at (-1.2,0);
      \coordinate (v) at (0,0);
      \coordinate (o) at (1.2,0);
      \coordinate (a) at (-1,0);
      \coordinate (b) at (1,0);
      \def\r{0.5}
      \coordinate (c) at (0,-0.5);
      \draw (i) -- (o);
      \draw[thick] (c) -- (a) node[midway, sloped, allow upside down]{$>$};
      \draw[thick] (c) -- (b);
      \draw (c) ++(0,-\r) circle (\r);
    \end{tikzpicture}%
    \caption*{(b) $P_{1\mathrm{vx}}$}
  \end{minipage}
    \hspace{0.04\linewidth}
   \begin{minipage}[c]{0.25\linewidth}
    \centering
    \begin{tikzpicture}[line width=1pt,baseline=(v.base),scale=0.7]
      \coordinate (i) at (-1.2,0);
      \coordinate (v) at (0,0);
      \coordinate (o) at (1.2,0);
      \coordinate (a) at (-1,0);
      \coordinate (b) at (1,0);
      \def\r{0.4}
      \coordinate (c) at (0,-0.4);
      \coordinate (d) at (0,-0.9);
      \draw (i) -- (o);
      \draw[thick] (c) -- (a) node[midway, sloped, allow upside down]{$>$};
      \draw[thick] (d) -- (c) node[midway, sloped, allow upside down]{$>$};
      \draw[thick] (c) -- (b);
      \draw (d) ++(0,-\r) circle (\r);
    \end{tikzpicture}%
    \caption*{(c) $P_{\mathrm{tad}}$}
    \label{fig:tadpole}
  \end{minipage}%
  \caption{Three diagrams with first type UV divergence: (a) one-point function, (b) two-point function with one vertex and (c) tadpole diagram.}
  \label{fig:fig1}
\end{figure}\\
These diagrams are grouped together because they share the same UV-divergent behavior in their momentum integrals. (See Appendix~\ref{A} for our notations of these loop integrals, also extensively discussed in~\cite{Inomata:2025bqw,Inomata:2025pqa,Kristiano:2025ajj}).

Finally, we consider the double-vertex contributions to the two-point function. Based on their propagator structure, this part can be decomposed into two components, $P_{2\mathrm{vx},a}$ and $P_{2\mathrm{vx},b}$, corresponding to the diagrams shown in Fig.~\ref{fig:fig2}(a) and (b), respectively. As we will demonstrate in the next section, these two diagrams exhibit different UV behaviors due to the coupling between their time and momentum integrations.
\begin{figure}[ht]
  \centering
  \begin{minipage}[c]{0.25\linewidth}
    \centering
    \begin{tikzpicture}[line width=1pt,baseline=(v.base),scale=0.7]
      \coordinate (i) at (-1.2,0);
      \coordinate (v) at (0,0);
      \coordinate (o) at (1.2,0);
      \coordinate (a) at (-1,0);
      \coordinate (b) at (1,0);
      \def\r{0.5}
      \coordinate (c) at (0,-1);
      \draw (i) -- (o);
      \draw (c) circle (\r);
      \coordinate (f) at ({-\r / sqrt(2)}, {-1 + \r / sqrt(2)});
      \coordinate (g) at ({\r}, {-1});
      \draw[thick] (a) -- (f) node[midway, sloped, allow upside down]{$<$};
      \draw[thick] (b) -- (g);
      \draw[thick] (c) ++(-\r,0) arc (180:270:\r) node[midway,sloped,allow upside down]{$<$};
    \end{tikzpicture}%
    \caption*{(a) $P_{2\mathrm{vx},a}$}
  \end{minipage}%
  \hspace{0.04\linewidth}
  \begin{minipage}[c]{0.25\linewidth}
    \centering
     \begin{tikzpicture}[line width=1pt,baseline=(v.base),scale=0.7]
      \coordinate (i) at (-1.2,0);
      \coordinate (v) at (0,0);
      \coordinate (o) at (1.2,0);
      \coordinate (a) at (-1,0);
      \coordinate (b) at (1,0);
      \def\r{0.5}
      \coordinate (c) at (-\r,-1);
      \coordinate (d) at ( \r,-1);
      \draw (i) -- (o);
      \draw[thick] (c) -- (a) node[midway, sloped, allow upside down]{$>$};
      \draw[thick] (d) -- (b) node[midway, sloped, allow upside down]{$>$};
      \draw (v) ++(0,-1) circle (\r);
    \end{tikzpicture}%
    \caption*{(b) $P_{2\mathrm{vx},b}$}
  \end{minipage}

  \caption{Two diagrams with different type of UV divergence}
  \label{fig:fig2}
\end{figure}

\section{Divergences and Counterterms}
In this section, we systematically analyze the UV divergences of each aforementioned loop diagram. When evaluating loop corrections, we integrate over the 3-momentum $\mathbf{k}$. In the UV region ($k \rightarrow +\infty$), The integrand contains deep horizon modes. For non-attractor inflation, the behavior of these modes may change, thereby altering the structure of the divergence.

First, we analyze the divergence of the loop diagrams shown in Fig.~\ref{fig:fig1}. Extracting their momentum integration parts yields a common integral structure:
\begin{equation}
D_1 = \int \frac{\mathrm{d}^3 \mathbf{k}}{(2\pi)^3} |u_{k}(\eta')|^2.
\end{equation}
Analyzing the UV behavior of this integral translates to analyzing the UV asymptotic form of the mode functions~\cite{Braglia:2026fle,Ballesteros:2025nhz}. Defining $W_k(\eta) \equiv a(\eta)u_k(\eta)$, whose evolution obey the Mukhanov-Sasaki equation $\frac{\ddd^2}{\ddd \eta^2}W_k + \omega_k^2(\eta)W_k = 0$, where the effective mass $\omega_k(\eta)$ is:
\begin{equation}
\omega_k^2(\eta) \equiv k^2 - 2a^2H^2 + a^2V^{(2)} \equiv k^2 \left[ 1 + \frac{M(\eta)}{k^2} \right].
\end{equation}
In the deep UV limit ($k \rightarrow +\infty$), $\omega_k$ is dominated by $k$, implying $\omega_k' \ll \omega_k$. Following the adiabatic approximation~\cite{Wentzel:1926aor,Kramers:1926njj}, we adopt the ansatz $W_k \simeq (2E_k)^{-1/2}\exp[-i\int^\eta E_k(\eta')\ddd\eta']$. The effective frequency $E_k$ can be expanded as
\begin{equation}
E_k^2 \approx \omega_k^2 + \frac{3\omega_k'^2}{4\omega_k^2} - \frac{\omega_k''}{2\omega_k} = k^2 \left[ 1 + \frac{M}{k^2} + O(k^{-4}) \right].
\end{equation}
Utilizing this adiabatic expansion, we separate the divergent terms from the finite parts in $D_1$ to obtain the regularized result $R_1$:
\begin{equation}
\begin{aligned}
D_1 &= f_1 \int^{+\infty}\ddd k\, \frac{k^2}{E_k} = f_1\int^{+\infty}\ddd k\,\left[k - \frac{M}{2k}\right] + F_1\\
&\Rightarrow R_1 = \frac{f_1}{2}\Lambda^2 - \frac{Mf_1}{2}\ln{\frac{\Lambda}{k_*}} + F_1,
\end{aligned}
\end{equation}
where $f_1 \equiv 1/(2\pi a)^2$ and $F_1$ denotes the finite part. Notably, this integral contains both quadratic and logarithmic divergences. The logarithmic divergent term proportional to $V^{(2)}$ differs from standard SR inflation results. Here, we maintain a general regularization form parameterized by an energy scale $\Lambda(\eta)$ and $k_*(\eta)$, where tuning $k_*$ shifts the boundary between the logarithmic and finite terms.

For the double-vertex diagrams in Fig.~\ref{fig:fig2}, the loop integrals contain phase factors $e^{ik(\eta-\eta')}$ that couple the momentum and time integrations, requiring a more delicate analysis~\cite{Ballesteros:2025nhz}. In the external IR limit, the potentially divergent part is
\begin{equation}
D_2 = 2\int^{\eta'}\ddd\eta''a^4V^{(3)}_r\dot{\bar{\phi}}\int\frac{\ddd^3k}{(2\pi)^3}u_k^2(\eta')u_k^{*2}(\eta'').
\end{equation}
The integrand includes a phase factor $e^{-2ikt}$, where $t \equiv \eta'-\eta''$. To correctly account for the interaction-induced corrections to the vacuum state, we perform a slight Wick rotation in the time variable $\eta'' \rightarrow \eta''(1-i\theta)$. This yields an exponential suppression $e^{-2k\theta t}$ in the UV limit, rendering the integral finite whenever $t \neq 0$. Consequently, $D_2$ diverges only in the coincident limit $k \rightarrow +\infty$ and $t \rightarrow 0$. By integrating over time first, we extract the leading logarithmic divergence:
\begin{equation}\label{R2}
\begin{aligned}
D_2 \approx& f_2 \int^{+\infty}\ddd k \int_0^{t_0}\ddd t \, e^{-2ik t(1-i\theta)}+\dots\\
=& f_2 \int^{+\infty}\ddd k \frac{i}{2k}\left(e^{-2ik t_0(1-i\theta)}-1\right)+\dots\\
\Rightarrow& R_2 = \frac{-if_2}{2}\ln{\frac{\Lambda}{k_*}} + F_2,
\end{aligned}
\end{equation}
where $f_2 \equiv V_r^{(3)}\dot{\bar{\phi}}/4\pi^2$ and $F_2$ is finite. A detailed derivation is provided in Appendix~\ref{B}. Crucially, Fig.~\ref{fig:fig2}(a) corresponds to the imaginary part of $D_2$ and thus exhibits a logarithmic divergence. Conversely, Fig.~\ref{fig:fig2}(b) corresponds to the real part, which is purely finite and additionally suppressed by $q^3$ in the IR limit.

To cancel these divergences, we introduce specific counterterms. For the one-point function, a linear counterterm $A_1\delta\phi$ yields a contribution $\langle\delta\phi\rangle_c = -\int \mathrm{d}\eta' a^2 G_0 A_1$. By choosing $A_1 = -\frac{1}{2}V_r^{(3)}R_1 + \tilde{F}_1$, we cancel the divergences in both the one-point function and the tadpole diagram (Fig.~\ref{fig:fig1}c). Since the macroscopic background $\bar{\phi}$ is typically defined to absorb all backreaction corrections (i.e., setting $\langle\delta\phi\rangle = 0$), we simultaneously fix the finite part $\tilde{F}_1 = 0$.

In the superhorizon limit of the external momentum, the regularized one-loop contribution to the power spectrum takes the form:
\begin{equation}
\begin{aligned}
P_{\mathrm{loop}} =& -\frac{q^3}{2\pi^2} \int^{\eta} \mathrm{d}\eta' a^2 G_q(\eta,\eta') \mathrm{Re}\left[ u_{q}(\eta) u_{q}^*(\eta') \right]\\
&\times \left( V_r^{(4)} R_1 + \frac{V_r^{(3)}(\eta')}{\dot{\bar{\phi}}(\eta')}\mathrm{Im}R_2 \right).
\end{aligned}
\end{equation}
To absorb this structure, we introduce a quadratic counterterm $A_2\delta\phi^2$. The full divergence is canceled by setting $A_2 = -\frac{1}{4}\left(V_r^{(4)}R_1 + \frac{V_r^{(3)}}{\dot{\bar{\phi}}}\mathrm{Im}R_2\right) + \tilde{F}_2$. This leaves a residual finite degree of freedom, $\tilde{F}_2$. Without imposing further physical constraints on the counterterms, the renormalized loop results remain ambiguous. The ultimate physical prediction depends entirely on how $\tilde{F}_2$ is fixed—a problem we will resolve in the next section by invoking the Ward identities associated with dilatation symmetry.

\section{Tuning counter terms by symmetry}
We can find an analogy in gauge field theory, results after renormalization will differ by a finite term when different regularization schemes are adopted. This arises because many regularization schemes, break gauge symmetry \cite{tHooft:1972tcz}. Adopting symmetry-preserving regularization schemes, such as dim-reg, is one class of solutions to this kind of problem \cite{Ema:2026dop}. However, this does not mean other regularization schemes cannot be used. For instance, when calculating the photon self-energy correction in QED, a quadratic divergence that can only be canceled by a mass term arises if a cutoff regularization scheme is adopted \cite{Peskin:1995ev}.
This counterterm appears to explicitly break the symmetry. However, if we then use the Ward–Takahashi identity to fix the finite part of this counterterm, the renormalized one-loop mass correction will ultimately vanish, giving a loop result consistent with the symmetry \cite{Kluberg-Stern:1974iel}. The situation is analogous for the problem studied in this paper. 

However, the issue is that gauge symmetry exists as a first principle of gauge field theory, so we must preserve the symmetry in all calculations. In inflationary field theory, by contrast, which symmetry can fix the loop order Lagrangian? A reasonable choice is to assume that the bare Lagrangian obeys the same symmetries as at tree level. Several works have pointed out that a certain dilatation symmetry is closely related to the IR conservation of the power spectrum. Considering such a transformation parameterized by an infinitesimal constant $\lambda$
\begin{equation}
\tilde{\boldsymbol{x}} = (1 - \lambda)\boldsymbol{x}, \,
\tilde{t} = t + \frac{\lambda}{H}, \, 
\delta\tilde{\phi}(\tilde{\boldsymbol{x}}, \tilde{t}) = \delta\phi(\boldsymbol{x}, t) - \lambda \frac{\dot{\bar{\phi}}}{H}\,.
\end{equation}
we can verify that the tree-level action is invariant under this transformation. 

Starting from this symmetry, we wish to first determine the conditions that a symmetry preserving regularization scheme must satisfy. To investigate this problem, we start from the corresponding Ward identity. Here, we are going to briefly introduce the derivation and leaving all details in the Appendix \ref{C}.

It is well known that the most standard derivation of the Ward identity consists of performing a change of integration variables on the correlation functions expressed in path-integral form.
After such a substitution, the correlation function often acquires additional terms formally, whereas the variable change itself leaves the integral unchanged.
Consequently, these formal extra terms must vanish, yielding an identity.
If this variable transformation leaves the action invariant, this gives precisely the Ward identity. In our case, the field transformation $\delta\phi\rightarrow \Psi$ can be defined such that $|\Psi\rangle\equiv|\delta\phi((1+\lambda)\mathbf{x})-\lambda\frac{\dot{\bar{\phi}}}{H},t+\frac{\lambda}{H}\rangle$. Since the action is invariant under this transformation, the relation $\langle\Psi_1|\Psi_2\rangle=\langle\delta\phi_1|\delta\phi_2\rangle$ holds for the transition matrix element. Now, we expand a two-point correlation function in the eigenstates of the field configuration and change the variable
\begin{equation}\label{path}
\begin{aligned}
&\langle\Omega|\delta\hat{\phi}(\mathbf{x})\delta\hat{\phi}(\mathbf{y})|\Omega\rangle
=\int\mathrm{D}\delta\phi_1\mathrm{D}\delta\phi_2\mathrm{D}\delta\phi_{-}\mathrm{D}\delta\phi_{+}\delta\phi_1\delta\phi_2\\
&\times\langle\Omega|\delta\phi_{-}\rangle\langle\delta\phi_{-}|\delta\phi_1\rangle\langle\delta\phi_1|\delta\phi_2\rangle\langle\delta\phi_2|\delta\phi_{+}\rangle\langle\delta\phi_{+}|\Omega\rangle,\\
&=\int\mathrm{D}\Psi_1\mathrm{D}\Psi_2\mathrm{D}\Psi_{-}\mathrm{D}\Psi_{+}\times\langle\Omega|\delta\phi_{-}\rangle\langle\Psi_{-}|\Psi_1\rangle\\
&\times\langle\Psi_1|\Psi_2\rangle\langle\Psi_2|\Psi_+\rangle\langle\delta\phi_{+}|\Omega\rangle\,\delta\phi_1\delta\phi_2,
\end{aligned}
\end{equation}
where $|\delta\phi_{-}\rangle$ and $|\delta\phi_{-}\rangle$ are eigenstates at an early time $t_i$, when the interactions have not yet been activated. Thus, the vacuum wave function in this case contains no complicated non-Gaussian terms, and we may safely consider the BD vacuum wave function \cite{Hui:2018cag,Maldacena:2002vr,Chen:2017ryl}. We can prove the following transformation relation of the wave function \cite{Hinterbichler:2013dpa}
\begin{equation}
\langle\Omega|\Psi\rangle 
=\left(1+\lambda\epsilon_0 \frac{\dot{\bar{\phi}}}{H} \delta\phi_0\right)\langle \Omega | \delta \phi ,t_i\rangle\,.
\end{equation}
Substituting this transformation relation into Eq.\ref{path}, and then expanding all terms in the equation with respect to $\lambda$, we extract the new terms generated by the integral transformation. Setting these terms to zero yields an identity relating the squeezed limit three-point and two-point correlation functions \cite{Assassi:2012zq}
\begin{equation}\label{Ward3}
-\epsilon_0\frac{\dot{\bar{\phi}}_i}{H}[\langle\delta\hat{\phi}_{i0}\delta\hat{\phi}_1\delta\hat{\phi}_2\rangle+c.c.]=\langle\delta\delta\hat{\phi}_1\delta\hat{\phi}_2\rangle+1\leftrightarrow 2,
\end{equation}
where we defined $\delta\delta\hat{\phi}\equiv{x^i \partial_i \hat{\phi}} - 
\delta\dot{\hat{\phi}}/{H}-\dot{\bar{\phi}}/{H}$. 
By performing a perturbative expansion on Eq.\ref{Ward3}, while setting $t_1= t_2$ and $\mathbf{y}= \mathbf{x}$, we can extract the leading-order contributions that yield nontrivial relations after some calculations, which is precisely the integral identity
\begin{equation}
\dot{D_1}=\mathrm{Im}D_2.
\end{equation}
We have to emphasize that, we have used the property of zero modes here, which means there are no growing modes in its mode function. As a result, for finite momentum with growing modes, it has to be analyzed separately. To clarify its relation to regularization schemes, we first start from a differential identity: 
\begin{equation}\label{consis}
\begin{aligned}
&\frac{\mathrm{d} u_{\mathbf{k}}(t)}{\mathrm{d} t} =- H k^{-3/2} \frac{\mathrm{d}\left( k^{3/2} u_{\mathbf{k}}(\eta) \right)}{\mathrm{d} \ln k}\\
&+ 2 \int_{\eta_i}^{\eta} \mathrm{d}\eta' \, a^4 V_{(3)} \dot{\bar{\phi}}\,\mathrm{Im}\left[ u_{\mathbf{k}}(\eta) u_{\mathbf{k}}^*(\eta') \right] u_{\mathbf{k}}(\eta').
\end{aligned}
\end{equation}
This identity follows from the equation of motion satisfied by the mode functions of the scalar field \cite{Inomata:2024lud,Inomata:2025bqw,Maldacena:2002vr}. Using Eq.\ref{consis}, we can obtain 
\begin{equation}
\begin{aligned}
&\mathrm{Im}D_2=\int  \frac{\mathrm{d}^3 \mathbf{k}}{(2\pi)^3}\frac{\mathrm{d}}{\mathrm{d}t}|u_k|^2+\int\ddd k\frac{H}{2\pi^2}\frac{\ddd(k^3|u_k|^2)}{\ddd k}\\
\Rightarrow& \mathrm{Im}R_2=[\frac{\dot{f}_1}{2}\Lambda^2-\frac{f_2}{2}\ln{\frac{\Lambda}{k_*}}]+[-\frac{\dot{f}_1}{2}\Lambda^2]+F_2,
\end{aligned}
\end{equation}
where we have used the property $(f_1M)^\cdot=f_2$. It is not surprising that this result is consistent with Eq.\ref{R2}. For a symmetry-preserving regularization scheme, we should require that $\dot{R_1}=\mathrm{Im}R_2$, which necessitates taking the time derivative of the regularized result
\begin{equation}
\dot{R_1}=\frac{\dot{f}_1}{2}\Lambda^2+f_1\Lambda\dot{\Lambda}-\frac{f_2}{2}\ln{\frac{\Lambda}{k_*}}-\frac{Mf_1}{2}(\frac{\dot{\Lambda}}{\Lambda}-\frac{\dot{k}_*}{k_*})+\dot{F_1}.
\end{equation}
By equating the divergent terms and finite terms at each order, we obtain the differential equation satisfied by the regularization scale, as well as constraints on the finite parts
\begin{equation}
\dot{\Lambda}=H\Lambda,\quad F_2=\dot{F}_1-\frac{Mf_1}{2}(H-\frac{\dot{k_*}}{k_*}),
\end{equation}
where we have used $\dot{f}_1=-2Hf_1$. Properly defined physical momentum cutoff and dim-reg can fulfill these conditions, and we therefore identify them as symmetry-preserving schemes. We will show more examples in appendix \ref{D}.

Symmetry can constrain not only the regularization scheme but also the choice of counterterms. The symmetry considered in this work is, in fact, equivalent to assuming that all counterterms can be combined into a single counterterm potential $V_c(\bar{\phi})$, so the coefficients of the counterterms must satisfy the relation $\dot{A}_1=2\dot{\bar{\phi}}A_2$ (proved in appendix \ref{C}). 
Combined with the constraints on regularization derived earlier, this directly implies that the finite parts $\tilde{F}_2$ in the counterterm coefficients $A_2$ are exactly zero. This means that the symmetry essentially fix the IR limit of the power spectrum. 

Now, we have shown how symmetry constrains regularization and counterterms during loop calculations, thereby constraining one-loop corrections. However, this conclusion can be reached without any perturbative computation.
When we derived Eq.\ref{Ward3} previously, we considered identical transformations of the two-point correlation function. In fact, one can equally well consider transformations of the one-point correlation function \cite{Assassi:2012et}, which ultimately leads to (see appendix \ref{C})
\begin{equation}
-\epsilon_0\frac{\dot{\bar{\phi}}_i}{H}[\langle\delta\hat{\phi}_{i0}\delta\hat{\phi}\rangle+c.c.]=\langle\delta\delta\hat{\phi}\rangle.
\end{equation}
This identity directly provides the time evolution of an unequal-time correlation function. We note that for mode functions with \(q=0\), which contain no growing modes, their tree-level results already satisfy the Ward identities exactly, so loop corrections won't arise. 
This shows that the power spectrum evolution in the superhorizon limit is a nonperturbative property directly enforced by symmetry, independent of any specific loop calculation.
The present situation is completely analogous to the case of QED corrections to the electron mass. A symmetry-violating regularization may necessitate symmetry-violating counterterms, whereas the Ward identity serves to fix the coefficients of the counterterms.
For any regularization scheme, symmetry requires that the finite part 
$\tilde{F}_2$ in the counterterm coefficient $A_2$ vanishes, which would in turn lead to 
\begin{equation}
\dot{A}_1=2\dot{\bar{\phi}}A_2-(\dot{f}_1\Lambda^2+2f_1\Lambda\dot{\Lambda})+[F_b+\frac{Mf_1}{2}(\frac{\dot{\Lambda}}{\Lambda}-\frac{\dot{k}_*}{k_*})],
\end{equation}
forcing the counterterm to explicitly break the symmetry.

\section{conclusion and discussion}
From the derivation above, we can conclude that the loop corrections to the correlation functions depend strongly on the choice of regularization scheme and the finite coefficients in the counterterms \cite{Ballesteros:2025nhz}. The selection of these counterterms inherently relies on our assumptions regarding the underlying UV theory. By introducing the aforementioned symmetry and the Ward identity derived from it, we can simultaneously constrain both the regularization scheme and the counterterm coefficients, which is consistent with the findings in \cite{Ema:2026dop}. We further point out that this symmetry directly leads to an exact cancellation among loop diagrams, manifesting as a non-perturbative property that is independent of specific perturbative loop calculations.

It is worth clarifying where our Ward identity method provides crucial information beyond traditional tree-level consistency relations. Many works have pointed out that, by virtue of consistency relations, loop corrections in the IR limit of external momentum can be rearranged into an integral of total derivative terms \cite{Inomata:2024lud,Inomata:2025bqw,Inomata:2025pqa,Tada:2023rgp,Fumagalli:2024jzz,Kawaguchi:2024rsv,Fang:2025vhi}. However, the handling of the quadratic divergences originating from such boundary integrals is precisely the key issue leading to renormalization ambiguities. In contrast, our Ward identity approach starts from coordinate space and directly yields integral identities, thereby imposing strict extra constraints on the regularization procedure itself.

However, we must emphasize that this symmetry originates from the extension of the tree-level action symmetry, which intrinsically serves as an additional assumption about the UV theory. At present, observations heavily favor the tree-level predictions of canonical single-field inflation, but this does not guarantee that the theory remains fully describable by a single-field model once loop corrections are included. Therefore, the breaking of this symmetry would be a strong indicator of new physics, potentially serving as a cosmological collider \cite{Arkani-Hamed:2015bza,Lee:2016vti,Chen:2016nrs,Wang:2019gbi}. In other words, rather than relying on ambiguous perturbative loop computations, our approach demonstrates that the underlying symmetry fundamentally dictates the physical outcome.

Nevertheless, several important open issues remain beyond the scope of this work. First, our proof essentially considers loop diagrams with zero external momentum. Since such modes remain permanently superhorizon, their mode functions contain no growing modes. For modes with finite observable momentum, however, the presence of growing modes can generate non-trivial loop corrections at superhorizon scales. We plan to extend our calculations to finite-momentum models and address this issue in detail in a subsequent paper. Second, it is necessary to incorporate nonlinear gauge transformations to prove that the symmetry employed in this work is mathematically equivalent to the dilation symmetry of the comoving gauge \cite{Hinterbichler:2012nm,Assassi:2012et}. This remains a key objective for our future research.

\textit{\textbf{Acknowledgements.}} 
We thank Jason Kristiano, Chao Chen, Zhong-Zhi Xianyu, Yu-Ming Wang, Lei-Hua Liu, Tomislav Prokopec and Junichi Yokoyama for useful comments. This work is supported in part by the National Natural Science Foundation of China (No. 12475067 and No. 12235019).

\bibliographystyle{apsrev4-1}
\bibliography{main}
\appendix
\section{one-loop contributions}\label{A}
We consider quadratic action $S_{2}$ as free theory
\begin{equation}
S_{2}=\int \ddd \eta \ddd^3 x\ a^2\left(\frac{1}{2} \delta \phi^{\prime 2}-\frac{1}{2}\left(\partial_i \delta \phi\right)^2\right)-\frac{1}{2} V_r^{(2)} \delta \phi^2,
\end{equation}
we can first solve out the evolution of field operators in the interaction picture
\begin{equation}
\delta \phi=\int \frac{\ddd^3 \mathbf{q}}{(2 \pi)^3} e^{i \mathbf{q} \cdot \mathbf{x}}\left(u_q \hat{a}_{\mathbf{q}}+u_q^* \hat{a}_{-\mathbf{q}}^{\dagger}\right).
\end{equation}
the equation of motion of the mode function $u_k$ is
\begin{equation}
\left[\frac{\ddd^2}{\ddd \eta^2}+2 \mathcal{H} \frac{\ddd}{\ddd\eta}+k^2+a^2V_r^{(2)}\right] u_k=0.
\end{equation}
We introduce Green’s function $G_q$ to 
simplify the expressions
\begin{equation}
\begin{aligned}
\left[\delta\phi_{\mathbf{p}}\left(\eta^{\prime}\right), \delta\phi_{\mathbf{q}}(\eta)\right]^\prime&=\left(u_q\left(\eta^{\prime}\right)u_q^{*}(\eta)-u_q^{*}\left(\eta^{\prime}\right) u_q(\eta)\right) \\
		&\equiv  \frac{i}{a(\eta^\prime)^2}G_q\left(\eta , \eta^{\prime}\right).
\end{aligned}
\end{equation}
After that, we can compute correlation functions using the in-in formalism
\begin{equation}
\langle Q(\eta) \rangle = \langle 0 | \left( T e^{-i \int_{\eta_i}^{\eta} d\eta' H_{\mathrm{int}}} \right)^\dagger Q^I(\eta) \left( T e^{-i \int_{\eta_i}^{\eta} d\eta' H_{\mathrm{int}}} \right) | 0 \rangle,
\end{equation}
where $H_{\mathrm{int}}$ are interaction Hamiltonian determined by nonlinear order action $S_{\delta\phi}$. We start from one-point correlation functions, the one-loop correction of it reads
\begin{equation}
\begin{aligned}
\langle\delta\phi\rangle&= \int^{\eta} \mathrm{d}\eta' \frac{-a^2}{2} G_0(\eta,\eta')V_r^{(3)}\int \frac{\mathrm{d}^3 \mathbf{k}}{(2\pi)^3} |u_k(\eta')|^2\\
&=\int^{\eta} \mathrm{d}\eta' \frac{-a^2}{2} G_0(\eta,\eta')V_r^{(3)}D_1.
\end{aligned}
\end{equation}
Then we begin to consider the power spectrum defined as
\begin{equation}
 (2\pi)^3\delta^{(3)}(\boldsymbol{k}+\boldsymbol{p})P(k) \equiv \frac{k^3}{2\pi^2}\langle\delta \phi_{\boldsymbol{p}}\delta \phi_{\boldsymbol{k}}\rangle\,.
\end{equation}
Up to one-loop order, there are four contributing terms that correspond to four Feynman diagrams shown in Fig.\ref{fig:fig1} and Fig.\ref{fig:fig2}, respectively.
\begin{equation}
\begin{aligned}
&P_{1\mathrm{vx}}=\frac{-q^3}{2\pi^2} \int^{\eta} \mathrm{d}\eta' a^2 G_q(\eta,\eta') \mathrm{Re}\left[ u_{q}(\eta) u_{q}^*(\eta') \right] \\
&\times\left( V_r^{(4)} \int \frac{\mathrm{d}^3 \mathbf{k}}{(2\pi)^3} |u_{k}(\eta')|^2\right)\\
=&\frac{-q^3}{2\pi^2} \int^{\eta} \mathrm{d}\eta' a^2 G_q(\eta,\eta') \mathrm{Re}\left[ u_{q}(\eta) u_{q}^*(\eta') \right]V_r^{(4)}D_1.
\end{aligned}
\end{equation}
\begin{equation}
\begin{aligned}
&P_{2\mathrm{vx},a}(q)=\frac{ q^3}{2\pi^2} \int^{\eta} \mathrm{d}\eta' \int^{\eta'} \mathrm{d}\eta'' a^2(\eta') a^2(\eta'')\\
&\times V_r^{(3)}(\eta') V_r^{(3)}(\eta'') G_q(\eta,\eta') \mathrm{Re}\left[ u_{q}(\eta) u_{q}^*(\eta'') \right] \\
& \int \frac{\mathrm{d}^3 \mathbf{k}}{(2\pi)^3}\left( G_k(\eta',\eta'') \mathrm{Re}\left[ u_{\tilde{k}}(\eta') u_{\tilde{k}}^*(\eta'') \right] + k \leftrightarrow \tilde{k} \right)\\
&q\rightarrow0=-\frac{q^3}{2\pi^2} \int^{\eta} \mathrm{d}\eta' a^2 G_q(\eta,\eta') \\
&\times\mathrm{Re}\left[ u_{q}(\eta) u_{q}^*(\eta') \right] \left(\frac{V_r^{(3)}(\eta')}{\dot{\bar{\phi}}(\eta')}\mathrm{Im}D_2\right),
\end{aligned} 
\end{equation}
\begin{equation}
\begin{aligned}
&P_{2\mathrm{vx},b}(q)= \frac{ q^3}{2\pi^2} \int^{\eta} \mathrm{d}\eta' \int^{\eta'} \mathrm{d}\eta'' a^2(\eta') a^2(\eta'')\\
& \times V_r^{(3)}(\eta') V_r^{(3)}(\eta'') G_q(\eta,\eta')G_q(\eta,\eta'') \\
&\int \frac{\mathrm{d}^3 \mathbf{k}}{(2\pi)^3} \mathrm{Re}\left[ u_{k}(\eta') u_{k}^*(\eta'') u_{\tilde{k}}(\eta') u_{\tilde{k}}^*(\eta'') \right]\\
&q\rightarrow0=-\frac{q^3}{2\pi^2} \int^{\eta} \mathrm{d}\eta' a^2 G_q(\eta,\eta')\\
&\times G_q(\eta,\eta') \left(\frac{V_r^{(3)}(\eta')}{\dot{\bar{\phi}}(\eta')}\mathrm{Re}D_2\right),
\end{aligned} 
\end{equation}
\begin{equation}
\begin{aligned}
&P_{\mathrm{tad}}(q, \eta)=\frac{q^{3}}{\pi^{2}} \int^{\eta} \mathrm{d} \eta^{\prime} a^{4}\left(\eta^{\prime}\right) V_r^{(3)}\left(\eta^{\prime}\right)\\
&\times \mathrm{Im}\left[u_{q}(\eta)^2u_{q}^{*}\left(\eta^{\prime}\right)^{2}\right]\left\langle\delta \phi\left(\mathbf{x}, \eta^{\prime}\right)\right\rangle.
\end{aligned}
\end{equation}
where we have defined $\tilde{k}\equiv|\mathbf{q-\mathbf{k}}|$. 

In the standard Schwinger-Keldysh (in-in) formalism, perturbative expansions are conventionally expressed using the $+/-$ branch propagators. However, to systematically isolate the IR behavior of loop integrals, it is highly advantageous to reorganize these propagators into commutators and anti-commutators, graphically denoted as arrowed (a-lines) and non-arrowed (na-lines) propagators, respectively. 

The \textbf{a-line} represents the retarded Green's function, capturing the causal linear response of the perturbations. Crucially, it remains regular in the soft momentum limit ($k \to 0$), typically suppressed by a factor of $k^3$. Conversely, the \textbf{na-line} corresponds to the symmetric Wightman function, representing the intrinsic quantum fluctuations of the vacuum. Unlike the a-line, it exhibits a severe IR divergence on superhorizon scales, scaling as $k^{-3}$. This strict separation of IR properties is crucial for defining generalized one-particle-irreducible (1PI) diagrams and tracking the flow of IR divergences in multi-loop computations.
\label{appendix:calcu}
\section{adiabatic expansion}\label{B}
We know that if we define $W_k(\eta)\equiv a(\eta)u_k(\eta)$, then the evolution equation for $W_k$ can be expressed in terms of the MS-equation
\begin{equation}
\frac{\ddd^2}{\ddd \eta^2}W_k+\omega_k^2(\eta)W_k=0,
\end{equation}
where the effective mass $\omega_k(\eta)$ is defined as 
\begin{equation}
\begin{aligned}
&\omega_k^2(\eta)\equiv k^2-2a^2H^2+a^2V^{(2)}\\=&k^2[1+\frac{1}{k^2}(a^2V^{(2)}-2a^2H^2)]\equiv k^2[1+\frac{M(\eta)}{k^2}]
\end{aligned}
\end{equation}
When $k\rightarrow +\infty$, $\omega_k$ is dominated by $k$, and thus $\omega_k' \ll\omega_k$.
To establish the exact form of the mode function without loss of generality, we parameterize it in a general polar form, $W_k(\eta) = r_k(\eta)e^{i\theta_k(\eta)}$, where $r_k$ and $\theta_k$ are strictly real functions. The canonical commutation relations of the quantum field enforce the Wronskian condition $W_k W_k'^* - W_k^* W_k' = i$. Substituting the polar parameterization into this condition yields $-2i r_k^2 \theta_k' = i$, which imposes an exact kinematic constraint between the amplitude and the phase: $\theta_k'(\eta) = -1/(2r_k^2(\eta))$. 

Since $r_k^2 > 0$, the phase derivative is strictly required to be negative ($\theta_k' < 0$). By defining a strictly positive effective frequency $E_k(\eta) \equiv -\theta_k'(\eta) > 0$, the amplitude is universally locked to $r_k(\eta) = 1/\sqrt{2E_k(\eta)}$. Consequently, instead of postulating an ad hoc linear combination, the canonical Wronskian strictly constrains the exact mode function to the polar form:
\begin{equation}
W_k(\eta) = \frac{1}{\sqrt{2 E_k(\eta)}} \exp\left( -i \int_{\eta_0}^\eta E_k(\eta') \mathrm{d}\eta' \right).
\end{equation}
Substituting this exact parameterization into the Mukhanov-Sasaki equation yields a non-linear differential equation for $E_k$:
\begin{equation}\label{adi}
E_k^2 = \omega_k^2 + \frac{3E_k'^2}{4E_k^2} - \frac{E_k''}{2E_k}.
\end{equation}
Generally, solutions to this equation contain high-frequency oscillations that invalidate any large-$k$ asymptotic expansion. However, the Bunch-Davies vacuum imposes the asymptotic boundary condition $\lim_{\eta \to -\infty} E_k = k$ with all derivatives vanishing. This physical requirement strictly eliminates oscillatory Bogoliubov mixing, uniquely selecting a smooth adiabatic tracking solution. Only for this non-oscillatory state is the adiabatic expansion mathematically justified. In the UV limit ($\omega_k' \ll \omega_k^2$), the derivative terms are heavily suppressed, allowing us to take $E_k \approx \omega_k$ as the zeroth-order approximation to iteratively expand $E_k$.

 Then substituting this zeroth-order approximation into Eq.\ref{adi} yields the first-order approximation
\begin{equation}
E_k^{2}\approx \omega_k^2+\frac{3\omega_k'^2}{4\omega_k^2}-\frac{\omega_k''}{2\omega_k}=k^2[1+\frac{M}{k^2}-\frac{M''}{4k^4}+O(\frac{1}{k^6})].
\end{equation}
Using this result to obtain the asymptotic behavior of the mode function, we can analyze the divergent behavior of the loop momentum integral. Starting with an analysis of $D_1$, it corresponds to 
\begin{equation}
\frac{1}{E_k}=\frac{1}{k}[1-\frac{M}{2}\frac{1}{k^2}+\frac{M''}{8}\frac{1}{k^4}+\frac{3M^2}{8}\frac{1}{k^4}+O(\frac{1}{k^6})],
\end{equation}
we can see that only the first two terms contribute to the divergence in $D_1$ 
\begin{equation}
\begin{aligned}
D_1=&\frac{1}{2\pi^2}\int\ddd k\,k^2\frac{W_k^2}{a^2}\\
=&\frac{1}{(2\pi a)^2}\int\ddd k\frac{k^2}{E_k}\\
=&f_1\int^{+\infty}\ddd k\,[k-\frac{M}{2}\frac{1}{k}]+F_1,
\end{aligned}
\end{equation}
where $f_1\equiv 1/(2\pi a)^2$ and $F_1$ are finite terms. Another divergent integral to be considered is \(D_2\). Due to the presence of the phase factor, the time and momentum integrals become coupled in the divergence analysis. We therefore adopt the approach of performing the time integral first, followed by the momentum integral.
\begin{equation}
\begin{aligned}
D_2&=\int\ddd k\,\frac{k^2}{\pi^2}\int^{\eta'}\ddd\eta''V_r^{(3)}\dot{\bar{\phi}}\frac{e^{-2i\int^{\eta'}_{\eta''}E_k(\eta_1)\ddd\eta_1}}{4E_k(\eta')E_k(\eta'')}\\
=&\int\ddd k\,\frac{k^2}{\pi^2}\frac{V_r^{(3)}\dot{\bar{\phi}}}{4E_k^2}\int^{\eta'}\ddd\eta''[1+O(\eta'-\eta'')]\\
\times&e^{-2iE_k(\eta')(\eta'-\eta'')+O((\eta'-\eta'')^2)}\\
\approx&\int\ddd k\,\frac{k^2}{\pi^2}\frac{V_r^{(3)}\dot{\bar{\phi}}}{4E_k^2}\int^{\eta'}\ddd\eta''e^{-2iE_k(\eta')(\eta'-\eta'')}\\
\approx&f_2\int^{+\infty}\ddd k\int_0^{t_0}\ddd t e^{-2ik t(1-i\theta)}+\dots\\
=&f_2\int^{+\infty}\ddd k\frac{i}{2k}(e^{-2ik t_0(1-i\theta)}-1)+\dots\\
=&-\frac{if_2}{2}\int^{+\infty}\ddd k\frac{1}{k}+F_2,
\end{aligned}
\end{equation}
where $f_2\equiv V^{(3)}\dot{\bar{\phi}}/4\pi^2$ and $F_2$ are finite
terms.

\section{Ward identities}\label{C}
Considering such a transformation
\begin{equation}
\tilde{\boldsymbol{x}} = (1 - \lambda)\boldsymbol{x}, \,
\tilde{t} = t + \frac{\lambda}{H}, \, 
\delta\tilde{\phi}(\tilde{\boldsymbol{x}}, \tilde{t}) = \delta\phi(\boldsymbol{x}, t) - \lambda \frac{\dot{\bar{\phi}}}{H}\,.
\end{equation}
First, we examine the transformation of the tree level action
\begin{equation}
\begin{aligned}
&S[\delta\phi]\\
&=\int_{t_i}^{t_e}\mathrm{d}t\mathrm{d}^3x \,a^3\left[\Bigl(\frac{1}{2}(\dot{\bar{\phi}}+\delta\dot{\phi})^2 - \frac{(\partial_i\delta\phi)^2}{2a^2}\Bigr)- V_r(\bar{\phi}+\delta\phi)\right]\\
&\approx \int_{\tilde{t}_i}^{\tilde{t}_e} \, d\tilde{t}\, d^3 \tilde{x}\, \tilde{a}^3 \biggl[\frac{1}{2}\biggl[-  \frac{(\tilde{\partial}_i \delta \tilde{\phi})^2}{\tilde{a}^2}\\
&\qquad+ \biggl( \dot{\bar{\phi}}(\tilde{t}) - \lambda \left(\frac{\dot{\bar{\phi}}}{H} \right)^{\cdot}+ \delta \dot{\tilde{\phi}}  + \lambda \left(\frac{\dot{\bar{\phi}}}{H} \right)^{\cdot}\biggr)^2 \biggr]\\
&\qquad - V_r\biggl( \bar{\phi}(\tilde{t}) - \lambda\frac{\dot{\bar{\phi}}}{H} + \delta \tilde{\phi} + \lambda\frac{\dot{\bar{\phi}}}{H}\biggr)\biggr],\\
&=\int_{\tilde{t}_i}^{\tilde{t}_e} \, d\tilde{t}\, d^3 \tilde{x}\, \tilde{a}^3 \biggl[\frac{1}{2}\biggl[-  \frac{(\tilde{\partial}_i \delta \tilde{\phi})^2}{\tilde{a}^2}+ \biggl( \dot{\bar{\phi}}(\tilde{t}) + \delta \dot{\tilde{\phi}} \biggr)^2 \biggr]\\
&\qquad - V_r\bigl( \bar{\phi}(\tilde{t})+ \delta \tilde{\phi}\bigr)\biggr]\,\\
&=S[\delta\tilde{\phi}],
\end{aligned}
\end{equation}
where we have taken the decoupling limit. This property means that the transformation mentioned above is a symmetry of the tree level action. In what follows, all our derivations assume that this symmetry will be protected under loop corrections. From the perspective of counterterms, this requirement is equivalent to assuming that the counterterm coefficients follow the same structure as the renormalized coupling constants. Consequently, all counterterms can be expressed via a potential function $V_c(\phi)$, leading to the identity $\dot{A}_1=2\dot{\bar{\phi}}A_2$ for the first and second-order coefficients.Then, we expand correlation functions in terms of the field configuration eigenbasis, thereby rewriting them in the form of a path integral
\begin{equation}\label{path}
\begin{aligned}
&\langle\Omega|\delta\hat{\phi}(\mathbf{x},t_1)\delta\hat{\phi}(\mathbf{y},t_2)|\Omega\rangle\\
=&\int\mathrm{D}\delta\phi_1\mathrm{D}\delta\phi_2
 \langle\Omega|\delta\phi_1,t_1\rangle\langle\delta\phi_1,t_1|\delta\phi_2,t_2\rangle\\
&\qquad\times\langle\delta\phi_2,t_2|\Omega\rangle\,\delta\phi_1(\mathbf{x})\delta\phi_2(\mathbf{y})\\
=&\int\mathrm{D}\delta\phi_1\mathrm{D}\delta\phi_2\mathrm{D}\delta\phi_{-}\mathrm{D}\delta\phi_{+}
 \langle\Omega|\delta\phi_{-}\rangle\langle\delta\phi_{-}|\delta\phi_1,t_1\rangle\\
&\qquad\times\langle\delta\phi_1,t_1|\delta\phi_2,t_2\rangle\langle\delta\phi_2,t_2|\delta\phi_{+}\rangle\\
&\qquad\times\langle\delta\phi_{+}|\Omega\rangle\,\delta\phi_1(\mathbf{x})\delta\phi_2(\mathbf{y}).
\end{aligned}
\end{equation}
In this formula, $|\delta\phi_{-}\rangle$ and $|\delta\phi_{-}\rangle$ are eigenstates at an early time $t_i$, when the interactions have not yet been activated. At this stage, we focus our attention on the transition matrix elements between field configuration eigenstates, which can be expressed further in the form of a path integral
\begin{equation}
\begin{aligned}
&\langle\delta\phi_1,t_1|\delta\phi_2,t_2\rangle
=\int_{\delta\phi(t_2)=\delta\phi_2}^{\delta\phi(t_1)=\delta\phi_1}
 \mathrm{D}\delta\phi\,e^{iS[\delta\phi]},\\
&=\int_{\delta\tilde{\phi}(\mathbf{\tilde{x}},\tilde{t}_2)
  =\delta\phi_2(\mathbf{x})-\lambda\frac{\dot{\bar{\phi}}_2}{H}}
 ^{\delta\tilde{\phi}(\mathbf{\tilde{x}},\tilde{t}_1)
  =\delta\phi_1(\mathbf{x})-\lambda\frac{\dot{\bar{\phi}}_1}{H}}
 \mathrm{D}\delta\tilde{\phi}\,e^{iS[\delta\tilde{\phi}]},\\
&=\Bigl\langle
 \delta\phi_1((1+\lambda)\mathbf{x})-\lambda\frac{\dot{\bar{\phi}}_1}{H},
 t_1+\frac{\lambda}{H}
\Big|\\
&\quad
 \delta\phi_2((1+\lambda)\mathbf{x})-\lambda\frac{\dot{\bar{\phi}}_2}{H},
 t_2+\frac{\lambda}{H}
\Bigr\rangle.
\end{aligned}
\end{equation}
Using this approach, we can find out how symmetries at the Lagrangian level constrain the time evolution of the system. Next, we replace all transition matrix elements in eq.\ref{path} with their forms after the symmetry transformation. Meanwhile, we adopt a concise representation by denoting the transformed eigenstates $|\delta\phi((1+\lambda)\mathbf{x})-\lambda\frac{\dot{\bar{\phi}}}{H},t+\frac{\lambda}{H}\rangle$ as $|\Psi\rangle$.
\begin{equation}\label{trans}
\begin{aligned}
&\langle\Omega|\delta\hat{\phi}(\mathbf{x},t_1)\delta\hat{\phi}(\mathbf{y},t_2)|\Omega\rangle\\
=&\int\mathrm{D}\delta\phi_1\mathrm{D}\delta\phi_2\mathrm{D}\delta\phi_{-}\mathrm{D}\delta\phi_{+}
 \langle\Omega|\delta\phi_{-}\rangle\langle\Psi_{-}|\Psi_1\rangle\\
&\times\langle\Psi_1|\Psi_2\rangle\langle\Psi_2|\Psi_+\rangle\langle\delta\phi_{+}|\Omega\rangle\,\delta\phi_1(\mathbf{x})\delta\phi_2(\mathbf{y})\\
=&\int\mathrm{D}\Psi_1\mathrm{D}\Psi_2\mathrm{D}\Psi_{-}\mathrm{D}\Psi_{+}
 \left(1-\lambda\epsilon_0\frac{\dot{\bar{\phi}}_i}{H}\delta\phi_{-0}\right)\\
&\times \langle\Omega|\Psi_{-}\rangle\langle\Psi_{-}|\Psi_1\rangle\langle\Psi_1|\Psi_2\rangle\langle\Psi_2|\Psi_+\rangle\\
&\times \left(1-\lambda\epsilon_0\frac{\dot{\bar{\phi}}_i}{H}\delta\phi_{+0}\right)
 \langle\Psi_{+}|\Omega\rangle\,\delta\phi_1(\mathbf{x})\delta\phi_2(\mathbf{y}).
\end{aligned}
\end{equation}
where we performed a change of integration variables. At the same time, we set the initial state of the system as the BD vacuum and utilized the transformation properties of the BD vacuum wave function
\begin{equation}
\begin{aligned}
&\langle\Omega|\Psi\rangle
\propto \exp\left[ -\frac{1}{2}\epsilon_0 \left(t+\frac{\lambda}{H}\right)\psi_0^2  \right] \\
&\times \exp\left[ \int_{k\neq0} \frac{\mathrm{d}^3k}{(2\pi)^3} \left(-\frac{1}{2} \epsilon_k\left(t+\frac{\lambda}{H}\right) \psi_{\mathbf{k}} \psi_{-\mathbf{k}}\right) \right] \\
&=\exp\left( -\frac{1}{2}\epsilon_0 \delta\phi_0^2 + \lambda\epsilon_0 \frac{\dot{\bar{\phi}}}{H} \delta\phi_0 \right) \\
&\times \exp\left[ \int_{\tilde{k}\neq0} \frac{\mathrm{d}^3\tilde{k}}{(2\pi)^3} \frac{-(1+3\lambda)^2(1-3\lambda)^2}{2}  \epsilon_{\tilde{k}}(t) \delta\phi_{\mathbf{\tilde{k}}} \delta\phi_{-\mathbf{\tilde{k}}} \right] \\
&=\left(1+\lambda\epsilon_0 \frac{\dot{\bar{\phi}}}{H} \delta\phi_0\right)\langle \Omega | \delta \phi ,t\rangle\,.
\end{aligned}
\end{equation}
According to the definition of state $|\Psi\rangle$, we know that it satisfies the eigen equations

\begin{equation}
\delta\hat{\phi}(\mathbf{x}(1-\lambda),t+\frac{\lambda}{H})|\Psi\rangle=(\delta\phi(\mathbf{x})-\lambda\frac{\dot{\bar{\phi}}}{H})|\Psi\rangle.
\end{equation}
From this relation, we obtain the linear transformation form of the operator acting on a complete set of basis vectors, allowing the operator to be spectrally decomposed in this basis
\begin{equation}
\delta\hat{\phi}(\mathbf{x}(1-\lambda),t+\frac{\lambda}{H})+\lambda\frac{\dot{\bar{\phi}}}{H}\hat{I}=\int\mathrm{D}\Psi\delta\phi(\mathbf{x})|\Psi\rangle\langle\Psi|.
\end{equation}
\begin{equation}
\delta\hat{\phi}_0=\int \mathrm{D}\Psi \delta\phi_0 |{\Psi\rangle \langle\Psi|}+O(\lambda)
\end{equation}

After decomposition, operators can be compared to the path integral formulation. Therefore, the integral over field configurations in \ref{trans} can actually be integrated out, allowing it to be recast into operator form
\begin{equation}
\begin{aligned}
&\langle\Omega|\delta\hat{\phi}(\mathbf{x},t_1)\delta\hat{\phi}(\mathbf{y},t_2)|\Omega\rangle\\
=&\int\mathrm{D}\Psi_{-}\mathrm{D}\Psi_{+}
\Bigl(1-\lambda\epsilon_0\frac{\dot{\bar{\phi}}_i}{H}\delta\phi_{-0}-\lambda\epsilon_0\frac{\dot{\bar{\phi}}_i}{H}\delta\phi_{+0}\Bigr)\\
\times&\langle\Omega|\Psi_{-}\rangle\langle\Psi_{-}|\Bigl[\delta\hat{\phi}\bigl(\mathbf{x}(1-\lambda),t_1+\frac{\lambda}{H}\bigr)+\lambda\frac{\dot{\bar{\phi}}}{H}\Bigr]\\
\times&\Bigl[\delta\hat{\phi}\bigl(\mathbf{y}(1-\lambda),t_2+\frac{\lambda}{H}\bigr)+\lambda\frac{\dot{\bar{\phi}}}{H}\Bigr]|\Psi_+\rangle
\langle\Psi_{+}|\Omega\rangle\\
=&\langle\Omega|\delta\hat{\phi}(\mathbf{x},t_1)\delta\hat{\phi}(\mathbf{y},t_2)|\Omega\rangle\\
+&\lambda\langle\Omega|\Bigl[\frac{1}{H}\delta\dot{\hat{\phi}}(\mathbf{x},t_1)-\mathbf{x}^i\partial_i\delta\hat{\phi}(\mathbf{x})+\frac{\dot{\bar{\phi}}_1}{H}\Bigr]\delta\hat{\phi}(\mathbf{y},t_2)|\Omega\rangle\\
+&1 \leftrightarrow 2-\lambda\Bigl[\langle\Omega|\epsilon_0\frac{\dot{\bar{\phi}}_i}{H}\delta\phi_{i0}\delta\hat{\phi}(\mathbf{x},t_1)\delta\hat{\phi}(\mathbf{y},t_2)|\Omega\rangle+\text{c.c.}\Bigr]
\end{aligned}
\end{equation}
As a result, the final form of the Ward identity reads:
\begin{equation}\label{Ward3}
\begin{aligned}
&-\epsilon_0\frac{\dot{\bar{\phi}}_i}{H}\bigl[\langle\delta\phi_{i0}\delta\hat{\phi}(\mathbf{x},t_1)\delta\hat{\phi}(\mathbf{y},t_2)\rangle+\text{c.c.}\bigr]\\
&=\langle\delta\delta\phi(\mathbf{x},t_1)\delta\hat{\phi}(\mathbf{y},t_2)\rangle+1\leftrightarrow 2,
\end{aligned}
\end{equation}

which is similar to those in flat spacetime. 

It is known that the squeezed limit of the three-point correlation function and the two-point correlation function satisfies consistency relations, which are essentially consequences of the Ward identities. By performing a perturbative expansion on \ref{Ward3}, while setting $t_1=t_2$ and $\mathbf{x}=\mathbf{y}$
\begin{equation}
\begin{aligned}
&\langle\delta\delta\hat{\phi}(\mathbf{x},t)\delta\hat{\phi}(\mathbf{x},t)\rangle+\langle\delta\hat{\phi}(\mathbf{x},t)\delta\delta\hat{\phi}(\mathbf{x},t)\rangle\\
&\approx\mathbf{x}^i\partial_i\langle\delta\hat{\phi}^2(\mathbf{x})\rangle-\frac{1}{H}\frac{\ddd}{\ddd t}\langle\delta\hat{\phi}^2(\mathbf{x})\rangle \\
&=-\frac{1}{H}\frac{\ddd}{\ddd t}\langle\delta\hat{\phi}^2(\mathbf{x})\rangle\\
&=-\frac{1}{H}\frac{\mathrm{d}}{\mathrm{d}t}\int \frac{\mathrm{d}^3 \mathbf{k}}{(2\pi)^3}|u_k|^2
\end{aligned}
\end{equation}
\begin{equation}
\begin{aligned}
&\epsilon_0\frac{\dot{\bar{\phi}}_i}{H}\bigl[\langle\delta\phi_{i0}\delta\hat{\phi}(\mathbf{x},t)\delta\hat{\phi}(\mathbf{x},t)\rangle+\text{c.c.}\bigr]\\
&\approx\epsilon_0\frac{\dot{\bar{\phi}}_i}{H}\biggl[\int \frac{\mathrm{d}^3 \mathbf{k}}{(2\pi)^3}\langle\delta\hat{\phi}_{i0}\delta\hat{\phi}_\mathbf{k}\delta\hat{\phi}_{-\mathbf{k}}\rangle +\text{c.c.}\biggr]\\
&=-\epsilon_0\frac{\dot{\bar{\phi}}_i}{H}4\int_{\tau_i}^{\tau} \mathrm{d}\tau' a^4(\tau')V^{(3)}(\bar{\phi})\int \frac{\mathrm{d}^3 \mathbf{k}}{(2\pi)^3}\\
&\mathrm{Im}[u_k(\tau)u_k^*(\tau')]\mathrm{Re}[u_0(\tau_i)u_0^*(\tau')]\mathrm{Re}[u_k(\tau)u_k^*(\tau')]\\
&=\frac{4}{H}\int_{\tau_i}^{\tau} \mathrm{d}\tau' a^4(\tau')V^{(3)}(\bar{\phi})\dot{\bar{\phi}}(\tau')\\
&\times\int \frac{\mathrm{d}^3 \mathbf{k}}{(2\pi)^3}\mathrm{Im}[u_k(\tau)u_k^*(\tau')]\mathrm{Re}[u_k(\tau)u_k^*(\tau')],
\end{aligned}
\end{equation}
where we have extracted the leading-order contributions that yield nontrivial relations. Meanwhile, we have also used the property of zero modes to simplify the expressions. Above, we considered the identical deformation of the two-point correlation function under a change of integration variables. We may equally well consider the transformation of the one-point correlation function, which will yield 
\begin{equation}
\begin{aligned}
&\langle\Omega|\delta\hat{\phi}(\mathbf{x},t)|\Omega\rangle=\int\mathrm{D}\Psi_{-}\mathrm{D}\Psi_{+}\langle\Omega|\Psi_{-}\rangle\langle\Psi_{+}|\Omega\rangle\\
&\times\Bigl(1-\lambda\epsilon_0\frac{\dot{\bar{\phi}}_i}{H}\delta\phi_{-0}-\lambda\epsilon_0\frac{\dot{\bar{\phi}}_i}{H}\delta\phi_{+0}\Bigr)\\
&\times\langle\Psi_{-}|\Bigl[\delta\hat{\phi}\bigl(\mathbf{x}(1-\lambda),t+\frac{\lambda}{H}\bigr)+\lambda\frac{\dot{\bar{\phi}}}{H}\Bigr]|\Psi_+\rangle
\\
&=\langle\Omega|\delta\hat{\phi}(\mathbf{x},t)|\Omega\rangle\\
&+\lambda\langle\Omega|\Bigl[\frac{1}{H}\delta\dot{\hat{\phi}}(\mathbf{x},t)-\mathbf{x}^i\partial_i\delta\hat{\phi}(\mathbf{x})+\frac{\dot{\bar{\phi}}}{H}\Bigr]|\Omega\rangle\\
&+\lambda\Bigl[\langle\Omega|\epsilon_0\frac{\dot{\bar{\phi}}_i}{H}\delta\phi_{i0}\delta\hat{\phi}(\mathbf{x},t)|\Omega\rangle+\text{c.c.}\Bigr].
\end{aligned}
\end{equation}
By extracting several additional terms, we obtain an identity that directly constrains the evolution of the two-point correlation function
\begin{equation}
-\epsilon_0\frac{\dot{\bar{\phi}}_i}{H}[\langle\delta\hat{\phi}_{i0}\delta\hat{\phi}\rangle+c.c.]=\langle\delta\delta\hat{\phi}\rangle=-\frac{\dot{\bar{\phi}}}{H}.
\end{equation}
In the expression, the Gaussian kernel of the early-time wavefunction reads $\epsilon_0=1/2|u_0|^2$, which ultimately allows us to derive
\begin{equation}
\dot{\bar{\phi}}_i[\langle\delta\hat{\phi}_{i0}\delta\hat{\phi}_0\rangle+c.c.]=2\dot{\bar{\phi}}\langle\delta\hat{\phi}_{0}\delta\hat{\phi}_0\rangle.
\end{equation}
This result directly shows the relationship between symmetry and conservations.

\section{regularization}\label{D}
In this section, we discuss the regularization of divergent integrals appearing in loop diagrams. Strictly speaking, these integrals contain both UV and IR divergences. UV divergences mainly arise from our lack of knowledge about the complete theory at high energy scales, namely the form of the underlying counterterms. IR divergences, by contrast, stem from the fact that the Universe is infinite in extent, while any observer can only access a finite local region, which gives rise to divergences \cite{Urakawa:2009my,Urakawa:2010it,Urakawa:2010kr}.

In this work, we focus on the UV divergent part, and simply take a physical IR momentum cutoff. We first review several commonly used regularization schemes, and then briefly address IR divergences in next section. For the regularization of UV divergences, we can always first isolate the divergent part. 
\begin{equation}
\begin{aligned}
&D_1=\int \frac{\mathrm{d}^3 \mathbf{k}}{(2\pi)^3} |u_{k}(\eta')|^2\equiv I_1+F_1\\
&=I_1+ \int \frac{\mathrm{d}^3 \mathbf{k}}{(2\pi)^3}\left[|u_{k}(\eta')|^2-\frac{1}{2a^2k}[1-\frac{M}{2}\frac{1}{k^2}]\right]\\
&\mathrm{Im}D_2=\int  \frac{\mathrm{d}^3 \mathbf{k}}{(2\pi)^3}\frac{\mathrm{d}}{\mathrm{d}t}|u_k|^2+\frac{H}{k^2}
\frac{\ddd(k^3|u_k|^2)}{\ddd k}\\
&=\int  \frac{\mathrm{d}^3 \mathbf{k}}{(2\pi)^3}\frac{\mathrm{d}}{\mathrm{d}t}\left[|u_{k}|^2-\frac{1}{2a^2k}[1-\frac{M}{2}\frac{1}{k^2}]\right]\\
&+I_2+\int\frac{\mathrm{d}^3 \mathbf{k}}{(2\pi)^3}\left[\frac{H}{k^2}
\frac{\ddd(k^3|u_k|^2)}{\ddd k}-\frac{H}{a^2k}\right]
\end{aligned}
\end{equation}
We may then consider how to regularize these divergent parts. A commonly used class of schemes is momentum cutoff, with which the regularized results can be written as
\begin{equation}
\begin{aligned}
&I_1=f_1\frac{\Lambda^2}{2}-f_1\frac{k_{IR}^2}{2}\frac{a^2}{a_i^2}-\frac{Mf_1}{2}\ln{\frac{\Lambda}{k_{IR}\frac{a}{a_i}}}\\
&I_2=-\frac{f_2}{2}\ln{\frac{\Lambda}{k_{IR}\frac{a}{a_i}}}
\end{aligned}
\end{equation}
To check whether the regularization respects the symmetry, we need to take the time derivative of \(R_1\) and compare it with the result for \(R_2\)
\begin{equation}
\begin{aligned}
&\dot{I}_1=\dot{f}_1\frac{\Lambda^2}{2}+f_1\Lambda\dot{\Lambda}-\dot{f}_1\frac{k_{IR}^2}{2}\frac{a^2}{a_i^2}-f_1Hk_{IR}^2\frac{a^2}{a_i^2}\\
&-\frac{f_2}{2}\ln{\frac{\Lambda}{k_{IR}\frac{a}{a_i}}}-\frac{Mf_1}{2}\left(\frac{\dot{\Lambda}}{\Lambda}-H\right)\\
&\dot{F}_1=\int  \frac{\mathrm{d}^3 \mathbf{k}}{(2\pi)^3}\frac{\mathrm{d}}{\mathrm{d}t}\left[|u_{k}|^2-\frac{1}{2a^2k}[1-\frac{M}{2}\frac{1}{k^2}]\right]\\
&-\frac{k_{IR^3}}{2\pi^2}\frac{a^2}{a_i^2}H|u_{k_{IR}\frac{a}{a_i}}|^2+f_1H[k_{IR}^2\frac{a^2}{a_i^2}-\frac{M}{2}]
\end{aligned}
\end{equation}
\begin{equation}
\begin{aligned}
&\mathrm{Im}R_2=\int  \frac{\mathrm{d}^3 \mathbf{k}}{(2\pi)^3}\frac{\mathrm{d}}{\mathrm{d}t}\left[|u_{k}|^2-\frac{1}{2a^2k}[1-\frac{M}{2}\frac{1}{k^2}]\right]\\
&-\dot{f}_1\frac{k_{IR}^2}{2}\frac{a^2}{a_i^2}-\frac{f_2}{2}\ln{\frac{\Lambda}{k_{IR}\frac{a}{a_i}}}\\
&-\frac{f_1HM}{2}-\frac{k_{IR^3}}{2\pi^2}\frac{a^2}{a_i^2}H|u_{k_{IR}\frac{a}{a_i}}|^2\\
\end{aligned}
\end{equation}
It can be seen that most of these terms are naturally equal. By equating the remaining terms, we obtain the constraint equations imposed by symmetry on the regularization
\begin{equation}
\begin{aligned}
\dot{f}_1\frac{\Lambda^2}{2}+f_1\Lambda\dot{\Lambda}=0,\\
\frac{Mf_1}{2}\left(\frac{\dot{\Lambda}}{\Lambda}-H\right)=0.
\end{aligned}
\end{equation}
We note that these two conditions can only be satisfied simultaneously if the cutoff momentum is a physical cutoff $\Lambda=k_{UV}a/a_i$.
Conversely, if the cutoff is taken to be the comoving momentum, not only will extra quadratic divergent terms appear, but the relations among the finite terms will also be broken. Consequently, one has to introduce symmetry-breaking counterterms to remove these divergences. 

The case of dim-reg is more subtle. For minimal subtraction dim-reg, quadratic divergences are discarded, so this part of the requirement is naturally satisfied. Therefore, we only need to consider the remaining logarithmic divergence.
The regularized result can be written as
\begin{equation}
\begin{aligned}
R_1&=\mu^\delta\int  \frac{\mathrm{d}^D \mathbf{k}}{(2\pi)^D}\frac{-M}{4a^2k^3}\\
&=\frac{-M}{4a^2}\frac{S_D\mu^\delta}{(2\pi)^D}\int^{+\infty}_{k_{IR}\frac{a}{a_i}}\frac{\ddd k}{k^{1+\delta}}  \\
&=-\frac{f_1M}{2}(2\sqrt{\pi}\mu)^\delta\frac{k_{IR}^{-\delta}}{\delta}\left(\frac{a}{a_i}\right)^{-\delta}\\
&=-\frac{f_1M}{2}\left(\frac{1}{\delta}+\ln{\frac{2\sqrt{\pi}\mu}{k_{IR}\frac{a}{a_i}}}+O(\delta)\right)
\end{aligned}
\end{equation}

\begin{equation}
\begin{aligned}
\mathrm{Im}R_2&=\mu^\delta\int  \frac{\mathrm{d}^D \mathbf{k}}{(2\pi)^D}\frac{-\pi^2f_2}{k^3}\\
&=-\frac{f_2}{2}(2\sqrt{\pi}\mu)^\delta\frac{k_{IR}^{-\delta}}{\delta}\left(\frac{a}{a_i}\right)^{-\delta}\\
&=-\frac{f_2}{2}\left(\frac{1}{\delta}+\ln{\frac{2\sqrt{\pi}\mu}{k_{IR}\frac{a}{a_i}}}+O(\delta)\right),
\end{aligned}
\end{equation}
where an energy scale is introduced to balance the dimensions, and the dimensional deviation is defined as $\delta\equiv3-D$. We have also used the area of the $D$ dimensional sphere $S_D=2\sqrt{\pi}^D/\Gamma(D/2)$. Then we have to evaluate the time derivative
\begin{equation}
\dot{R_1}=-\frac{f_2}{2}\left(\frac{1}{\delta}+\ln{\frac{2\sqrt{\pi}\mu}{k_{IR}\frac{a}{a_i}}}\right)-\frac{f_1M}{2}\left(\frac{\dot{\mu}}{\mu}-H\right),
\end{equation}
which means that if we want to protect the symmetry, $\dot{\mu}=H\mu$ is again a necessary condition. From this perspective, since symmetry also imposes constraints on the finite parts of the integrals, standard dim-reg of three-dimensional integrals does not necessarily satisfy the symmetry requirements. We must likewise fine-tune the time evolution of the energy scale $\mu$ starting from symmetry in order to meet these conditions.

\section{Regularization of IR Divergences via Local Observables}
In quantum electrodynamics, IR divergences are resolved by recognizing that soft photons are unobservable, necessitating a redefinition of observables based on physical cross-sections. A strictly analogous situation arises in cosmology: purely global quantities, such as the full-space average of $\delta\phi$ (which corresponds to the $k\rightarrow0$ limit), are unobservable to a local observer and naturally exhibit IR divergences. To regularize these, we must define genuine local observables. Consider a local observer in an FRW spacetime whose causal past intersects a spatial hyperplane, defining an observable region $\mathcal{O}$ centered at $x=0$. We introduce a window function $W_\eta(x)$ that is approximately unity within $\mathcal{O}$ and smoothly vanishes outside. Because an observer can only separate the background from perturbations using local information, the effective local background incorporates large-scale fluctuations:\begin{equation}\delta\bar{\phi} \equiv \hat{W}_\eta\delta\phi \equiv \frac{1}{V_\eta} \int\mathrm{d}^3 x \, W_\eta(x)\delta\phi,\end{equation}where $V_\eta\equiv\int\mathrm{d}^3 x \, W_\eta(x)$ is the normalization volume. To respect the expansion of the causal past, the window function must scale with the comoving time as $W_\eta(x) = W_{(1+\lambda)\eta}((1+\lambda)x)$.We then define the genuinely observable local perturbation as $\delta\tilde{\phi} \equiv \delta\phi - \delta\bar{\phi}$. While large-scale perturbations possess quantum uncertainty across the full ensemble of spatial patches, a local observer only ever measures a single patch. Consequently, superhorizon modes effectively classicalize, acting merely as a shift in the local background. By applying the local projection operator $\hat{W}_\eta$ to the full equation of motion and taking the expectation value (noting that spatial derivatives vanish under homogeneity and isotropy), we obtain the evolution for the local background:
\begin{equation}
\begin{aligned}
&\delta\bar{\phi}'' + 2\mathcal{H}\delta\bar{\phi}' + a^2V^{(2)}(\bar{\phi}+\delta\bar{\phi})\delta\bar{\phi} \\
&+ \frac{a^2}{2}V^{(3)}(\bar{\phi}+\delta\bar{\phi}) \hat{W}_\eta\langle\delta\tilde{\phi}^2(x)\rangle = 0.
\end{aligned}
\end{equation}
The projected two-point function takes the form:
\begin{equation}
\hat{W}_\eta\langle\delta\tilde{\phi}^2(x)\rangle = \int\frac{\mathrm{d}^3k}{(2\pi)^3} \langle\delta\phi_k\delta\phi_{-k}\rangle' \left(1 - \frac{|W_k|^2}{V_\eta^2}\right).\end{equation}Subtracting the background equation from the full Eom yields the dynamics for $\delta\tilde{\phi}$. Solving for the second-order correction using the Green's function, and restricting the integration to the observable region via $W_{\eta'}(x')$, we find:
\begin{equation}
\begin{aligned}
&\delta\tilde{\phi}(x) = \int^\eta\mathrm{d}\eta' \int\frac{\mathrm{d}^3k}{(2\pi)^3} G_k(\eta,\eta')e^{ikx} \frac{a^2}{2}V^{(3)}(\bar{\phi}+\delta\bar{\phi}) \\
&\int\frac{\mathrm{d}^3p}{(2\pi)^3} \left[\int\frac{\mathrm{d}^3p'}{(2\pi)^3}  \left(W_{\eta'}(p') - \frac{2}{V_{\eta'}}W_{\eta'}(p'-p)W_{\eta'}(p)\right) \right. \\
&\left.\delta\phi_p\delta\phi_{k-p-p'} - \langle\delta\phi_p\delta\phi_{-p}\rangle\left(1-\frac{|W_p|^2}{V_{\eta'}^2}\right)W_{\eta'}(k)\right].
\end{aligned}
\end{equation}
The structure $\left(1 - |W_p|^2/V_{\eta'}^2\right)$ explicitly cancels the IR divergences in the momentum integrals. For observable modes ($k$ much larger than the scale of $\mathcal{O}$), the second term in the bracket only affects the deep infrared and is negligible, leaving a finite local perturbation $\delta\tilde{\phi}_k$.To make this concrete, consider a window function mimicking a physical IR cutoff:\begin{equation}W_\eta(x)=\frac{-3}{k_{IR}^2x^2}\left(\cos(k_{IR}x)-\frac{1}{xk_{IR}}\sin(k_{IR}x)\right),\end{equation}where $k_{IR}(\eta)\equiv\Lambda_0\frac{a(\eta)}{a(\eta_0)}$. Since $\mathcal{H}\approx -1/\eta$ during inflation, this satisfies the required scaling $k_{IR}(\eta(1+\lambda))\approx (1-\lambda)k_{IR}(\eta)$. This demonstrates that a physical momentum cutoff naturally arises from defining genuine local observables, effectively regularizing IR modes. By contrast, a comoving cutoff fails to satisfy the causal scaling of the window function, underscoring why it yields unphysical results.

\end{document}